\documentclass[12pt]{article} 
\usepackage{graphicx} 

\def \bea{\begin{eqnarray}} 
\def \beq{\begin{equation}}

\def \eea{\end{eqnarray}} 
\def \eeq{\end{equation}}

\def \3half{\frac{3}{2}}

\def \s{\sqrt{2}}

\def \slk{k \! \! /}  
\def \slq{q \! \! /}

\begin{document} 
\begin{flushright} 
TECHNION-PH-2010-11 \\
EFI 10-20\\ 
August 2010 \\ 
arXiv:1008.4354 \\ 
\end{flushright} 
\centerline{\bf Second order direct CP asymmetry in $B_{(s)} \to X \ell \nu$} 
\bigskip 
\centerline{Shaouly Bar-Shalom, Gad Eilam, Michael Gronau} 
\medskip 
\centerline{\it Physics Department, Technion - Israel Institute of Technology} 
\centerline{\it Haifa 32000, Israel} 
\medskip 
\centerline{Jonathan L. Rosner} 
\medskip 
\centerline{\it Enrico Fermi Institute and Department of Physics, 
 University of Chicago} 
\centerline{\it Chicago, IL 60637, U.S.A.} 
\bigskip 
\begin{quote} 
A direct CP asymmetry in inclusive semileptonic $B_{(s)}$ decays 
vanishes by CPT to lowest order in weak interactions. Calculating 
the asymmetry at second order weak interactions in the 
Cabibbo-Kobayashi-Maskawa framework we find $A_{sl} = (-3.2 \pm 
0.9)\times 10^{-9}$. A maximal asymmetry which is two orders of 
magnitude larger is estimated in a left-right symmetric model. 
This quite generic upper bound implies negligible effects on 
wrong-sign lepton asymmetries in $B^0$ and $B_s$ decays. 
\end{quote} 
 
\leftline{\qquad PACS codes:  12.15.Hh, 12.15.Ji, 13.25.Hw, 14.40.Nd} 
 
\section{Introduction} 
 
The D0 Collaboration working at the Fermilab Tevatron has reported recently a 
charge asymmetry in like-sign dimuon events produced in $\bar p p$ collisions. 
The measured asymmetry~\cite{Abazov:2010hv,Abazov:2010hj}, 
\beq\label{D0} 
A^b_{sl} \equiv \frac{N^{++} - N^{--}}{N^{++} + N^{--}} = 
[-0.957 \pm 0.251~({\rm stat}) \pm 0.146~({\rm syst})]\%~, 
\eeq 
 was interpreted as due to CP violation in $B^0$-$\bar B^0$ or $B_s$-$\bar B_s$ 
mixing, 
\beq 
A^b_{sl} = (0.506 \pm 0.043)A^d_{sl} + (0.494 \pm 0.043)A^s_{sl}~. 
\eeq
(An asymmetry consistent with zero was measured a few years ago by 
the CDF collaboration using fewer same-sign dimuon events
produced in $\bar pp$ collisions with a lower integrated luminosity~\cite{CDF}.)
The experimental result (\ref{D0}) differs by $3.2$ standard deviations from 
much smaller asymmetries predicted within the 
Cabibbo-Kobayashi-Maskawa (CKM) framework~\cite{Lenz:2006hd}, 
\beq\label{predict} 
A^d_{sl} = (-4.8^{+1.0}_{-1.2}) \times 10^{-4}~,~~~~~~~~~ 
A^s_{sl} = (2.06 \pm 0.57) \times 10^{-5}~. 
\eeq 
Unambiguous evidence for physics beyond the Standard Model requires 
(a) showing that the measured asymmetry is, indeed, due to $B^0$ or $B_s$ 
semileptonic decays and not due to background processes~\cite{GR}, and 
(b) confirming an anomalously large negative asymmetry with a somewhat 
higher statistical significance than the current one. 
 
In view of the tiny CKM predictions (\ref{predict}) for $A^k_{sl}~(k=d, s)$ from 
second order $|\Delta B|=2$ weak transitions, attention has been often 
drawn to potentially larger corrections to the two asymmetries from new 
$|\Delta B|=2$ flavor physics at a TeV or higher energy 
scale~\cite{Browder:2008em}.  In this note we 
will study $|\Delta B| =1$  contributions to $A^k_{sl}$ from CP violation in 
inclusive semileptonic decays, which have been systematically neglected in 
all earlier studies. 
 
In Sec. II we review briefly the usual treatment of the asymmetry $A^k_{sl}$ 
involving CP violation in $B^0_k$-$\bar B^0_k$ mixing, introducing a new 
contribution from direct CP violation. Sec. III discusses an argument based 
on CPT for the vanishing of the new contribution at lowest order in 
weak interactions. We perform a second order calculation of the inclusive 
semileptonic asymmetry within the CKM framework. Maximal values of the 
asymmetry in a left-right extension of the Standard Model are studied 
in Sec. IV. In Sec. V we estimate for completeness second order amplitudes
for neutral $B$ mesons decaying directly into wrong-sign leptons while 
Sec. VI concludes. 
 
\section{Including a direct asymmetry in $A^k_{sl}$} 
One starts by defining neutral $B$ mass eigenstates $B^k_L$ and $B^k_H$, 
with mass and width differences $\Delta m_k$ and $\Delta\Gamma_k$,  in 
terms of flavor states $B^0_k$ and $\bar B^0_k$, 
\beq 
| B^k_L\rangle = p_k\,|B^0_k \rangle + q_k\,|\bar B^0_k \rangle~,~~~~~ 
| B^k_H\rangle = p_k\,|B^0_k \rangle - q_k\,|\bar B^0_k \rangle~. 
 \eeq 
Time evolution of flavor states~\cite{Branco:1999fs,Bigi:2000yz}, 
\bea 
|B^0_k(t) \rangle & = & g^k_+(t) |B^0_k \rangle - (q_k/p_k)\,g^k_-(t) |\bar B^0_k 
\rangle~,\nonumber\\ 
|\bar B^0_k(t) \rangle & = & g^k_+(t) |\bar B^0_k \rangle - (p_k/q_k)\,g^k_-(t) 
|B^0_k \rangle~, 
\eea 
implies time-dependent decay rates for inclusive semileptonic decays for 
wrong-sign leptons $B^0_k(t) \to X \ell^-\bar\nu_{\ell}$ ($\ell = e, \mu$) 
and their charge conjugates, 
\bea 
d\Gamma[B^0_k(t)\to X \ell^-\bar\nu_{\ell}]/dt & = & |\frac{q_k}{p_k}\bar 
A^k_{\ell}|^2\,|g^k_-(t)|^2~, \nonumber \\ 
\label{dGamma/dt} 
d\Gamma[\bar B^0_k(t)\to X \ell^+\nu_{\ell}]/dt & = & |\frac{p_k}{q_k} 
A^k_{\bar\ell}|^2\,|g^k_-(t)|^2~, 
\eea 
where 
\bea 
\bar A^k_{\ell} & \equiv & A(\bar B^0_k \to X \ell^-\bar\nu_{\ell})~,~~~ 
A^k_{\bar\ell} \equiv A(B^0_k \to X \ell^+ \nu_{\ell})~,\\ 
|g^k_-|^2 & = & \frac{1}{2}e^{-\Gamma_kt}[\cosh(\Delta\Gamma_kt/2) 
- \cos(\Delta m_kt)]~. 
\eea 
 
By definition $A^k_{sl}$ is the time-dependent asymmetry of wrong-sign leptons 
due to mixing, 
\beq\label{Asym} 
A^k_{sl}(t) \equiv \frac{d\Gamma[\bar B^0_k(t)\to X \ell^+\nu_{\ell}]/dt - 
d\Gamma[B^0_k(t) \to X \ell^-\bar\nu_{\ell}]/dt} 
{d\Gamma[\bar B^0_k(t)\to X \ell^+ \nu_{\ell}]/dt + d\Gamma[B^0_k(t) \to 
X \ell^- \bar\nu_{\ell}]/dt}~. 
\eeq 
Usually, one neglects a direct CP asymmetry in $B^0_k \to X \ell^+ \nu_{\ell}$. 
Assuming $|\bar A^k_{\ell}| = |A^k_{\bar\ell}|$ one 
finds~\cite{Branco:1999fs,Bigi:2000yz}, 
\beq\label{CPmix} 
A^k_{sl}({\rm mixing}) = \frac{1 - |q_k/p_k|^4}{1 + |q_k/p_k|^4} \approx {\rm Im} 
\left (\frac{\Gamma^k_{12}}{M^k_{12}}\right )~. 
\eeq 
That is, the asymmetry caused by CP violation in $B^0_k$-$\bar B^0_k$ mixing 
is given by ${\rm Im}(\Gamma^k_{12}/M^k_{12})$ where 
 $M^k_{12}$ and $\Gamma^k_{12}$ are off-diagonal elements of 
Hermitian matrices representing $B^0_k \leftrightarrow \bar B^0_k$ transitions 
via off-shell (dispersive) and on-shell (absorptive) intermediate states, 
respectively. The CKM predictions (\ref{predict}) for $A^k_{sl}({\rm mixing})$ 
are based on calculations of this imaginary part for $B^0$ and 
$B_s$~\cite{Lenz:2006hd}. 
 
We now define a direct CP asymmetry in inclusive semileptonic decays, 
\beq 
A^k_{sl}({\rm direct}) \equiv \frac{|\bar A^k_{\ell}|^2 - |A^k_{\bar\ell}|^2} 
{|\bar A^k_{\ell}|^2 + |A^k_{\bar\ell}|^2}~. 
\eeq 
Eqs.~(\ref{dGamma/dt}) and (\ref{Asym}) imply an expression for 
$A^k_{sl}(t)$ which includes both the 
asymmetry in mixing and the direct asymmetry.
Neglecting terms quadratic in the $A^k_{sl}({\rm mixing})$ and 
$A^k_{sl}({\rm direct})$, one has 
\beq 
A^k_{sl}(t) = \frac{A^k_{sl}({\rm mixing}) - A^k_{sl}({\rm direct})}
{1 - A^k_{sl}({\rm mixing})A^k_{sl}({\rm direct})} \approx
A^k_{sl}({\rm mixing}) - A^k_{sl}({\rm direct})~.
\eeq 
We note that this asymmetry of time-dependent decay rates is actually 
time-independent as in the special case (\ref{CPmix}) of 
CP violation in mixing alone.  In the following
two sections we will study $A^k_{sl}({\rm direct})$. 
 
\section{Direct asymmetry in the CKM framework} 
The inclusive semileptonic direct asymmetries in $B^0$ and $B_s$ decays 
are equal to each other to a good approximation as 
$\Gamma(B_s \to X \ell^+ \nu_{\ell}) = \Gamma(B^0 \to \ell^+ X \nu_{\ell}) + 
{\cal O}(m_s \Lambda_{\rm QCD}/m^2_b)$. (see also calculation 
below.) Furthermore, in the isospin symmetry limit the asymmetry in $B^0$ 
decays is equal  to that measured directly in self-tagged $B^+ \to X \ell^+\nu_{\ell}$ 
by comparing the rate for this inclusive process with that of its charge-conjugate. 
For this reason we omit the superscript $k$ in 
$A^k_{sl}({\rm direct})$ by defining a generalized direct semileptonic asymmetry 
for non-strange (charged or neutral) or strange $B$ mesons, 
\beq 
A_{sl} \equiv \frac{\Gamma(\bar B \to X \ell^- \bar\nu_{\ell}) - \Gamma(B \to X 
\ell^+ \nu_{\ell})}{\Gamma(\bar B \to X \ell^-\bar\nu_{\ell}) + \Gamma(B \to 
X \ell^+ \nu_{\ell})}~. 
\eeq 
 
It has often been stated that $A_{sl}$ vanishes because of CPT 
invariance~\cite{Branco_p58,Big_p186,Pais:1975qs}. 
CPT implies equal total decay widths for a particle and its antiparticle. 
A generalization 
of this theorem applies to partial decay rates for a set of final states, 
connected among themselves by strong and electromagnetic final 
state interactions but not connected by such interactions to other 
states. The inclusive decays $B \to X_{(C=-1)} \ell^+ \nu_{\ell}$ and 
$B\to X_{(C=0)} \ell^+ \nu_{\ell}$ are two special cases to which this 
generalization applies. 
 
A violation of the theorem of equal partial rates for $B \to X \ell^+ \nu_{\ell}$ 
and $\bar B \to X \ell^- \bar\nu_{\ell}$ is possible if one 
considers weak interactions which connect the final states $X \ell^+ \nu_{\ell}$ 
with intermediate hadronic states. That is, a small nonvanishing asymmetry 
$A_{sl}$ may be obtained by considering an interference between the 
dominant tree amplitude for $B \to X \ell^+\nu_{\ell}$ and an amplitude 
which is  second order in weak interactions. 
A similar interference has 
been shown to imply a tiny CP asymmetry in inclusive semileptonic rare 
top quark decays, $t \to d \ell^+\nu_{\ell}$~\cite{Soares:1992kr,Eilam:1992ks}. 
 
A very crude upper bound on this asymmetry is   
\beq\label{eqn:crude bound} 
|A_{sl}| < \frac{1}{4\pi}\frac{G_F}{\s}(m_b - m_c)^2 \sim {\rm few}\times 10^{-6}~. 
\eeq 
This estimate includes a suppression by a loop factor $1/4\pi$ and a phase
space  factor $(m_b -m_c)^2$.   Further suppression of the asymmetry may be due
to a small weak phase difference between the two interfering amplitudes and due
to a possible extra dynamical suppression of the second order amplitude. We will
show below that, indeed, such suppression factors exist in the CKM framework. 
The above upper limit is one order of magnitude smaller than the estimate  
(\ref{predict}) for a CP asymmetry in $B_s$-$\bar B_s$ mixing, and  
two orders of magnitude below the asymmetry in $B^0$-$\bar B^0$ mixing. 
  
\begin{figure}[h] 
\begin{center} 
\includegraphics[width=0.45\textwidth]{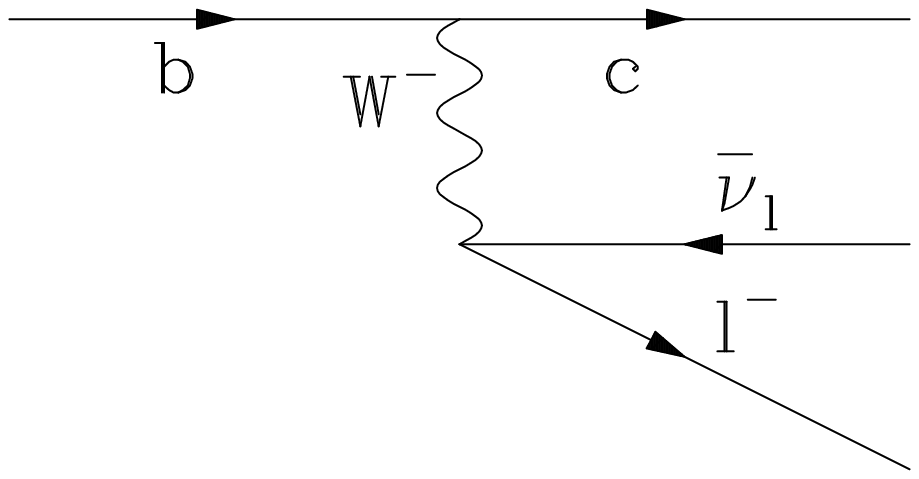} 
\end{center} 
\caption{Tree diagram for $b \to c \ell^- \bar\nu_{\ell}$ 
representing $\bar B \to X_{(C=1)} \ell^- \bar\nu_{\ell}$ 
\label{fig:tree}} 
\end{figure}

\begin{figure}[h] 
\begin{center} 
\includegraphics[width=0.55\textwidth]{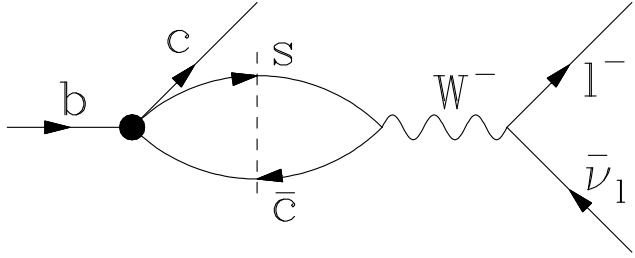} 
\end{center} 
\caption{Second order diagram for $b \to c \ell^- \bar\nu_{\ell}$ 
representing $\bar B \to X_{(C=1)} \ell^- \bar\nu_{\ell}$ 
\label{fig:loop}} 
\end{figure}

We will now calculate the CP asymmetry in CKM favored  
$B\to X_{(C=-1)}\ell^+\nu_{\ell}$ decays 
dominated by a tree amplitude proportional to $V^*_{cb}$ describing a 
quark transition $\bar b \to \bar c \ell^+\nu_{\ell}$.  In order to produce an 
asymmetry, the second amplitude, leading to the same final states, must 
involve a CKM factor with a different weak phase. A second order amplitude 
fulfilling these two requirements consists of a product of a penguin 
amplitude for $\bar b \to \bar c c\bar s$ involving $V^*_{tb}V_{ts}$ 
(see discussion below) and a tree amplitude for $c \bar s \to \ell^+ \nu_{\ell}$ 
involving $V^*_{cs}$. Two diagrams, describing the tree amplitude for 
$b \to c \ell^- \bar\nu_{\ell}$ and the second order amplitude for this 
transition, are shown in Figs.\ \ref{fig:tree} and \ref{fig:loop}, 
respectively. A relative CP-conserving phase of $90^\circ$ 
between the two amplitudes follows by taking the absorptive (i. e., imaginary) 
part of the second order amplitude. The absorptive part is described by a 
discontinuity cut 
crossing the $\bar c  s$ lines in the second order diagram, which amounts 
to summing over corresponding on-shell intermediate states. 

%
 
In order to calculate the asymmetry we write down expressions 
for an effective Hamiltonian associated with each of the three four-fermion 
vertices appearing in the two diagrams in Figs.\ \ref{fig:tree} and \ref{fig:loop}. 
The tree diagram is obtained from 
\beq\label{tree} 
H^{b\to c \ell^- \bar\nu_{\ell}}_{\rm eff} = \frac{G_F}{\s}V_{cb}[\bar c\gamma^{\mu} 
(1-\gamma_5)b][\bar\ell\gamma_{\mu}(1-\gamma_5)\nu_{\ell}] \label{m0}~. 
\eeq 
The Hamiltonian related to the first vertex in the second order diagram 
is~\cite{Buchalla:1995vs}, 
\bea\label{Heff} 
H^{b \to c \bar c s({\rm pen})}_{\rm eff} & = & \frac{G_F}{\s}V_{tb}V^*_{ts} 
(c_3O_3 + c_4O_4 + c_5O_5 + c_6 O_6)~,\\ 
c_3(m_b) & = &  0.012~,~~~~~O_3 =  [\bar s_\alpha\gamma^\mu(1-\gamma_5) 
b_\alpha][\bar c_\beta\gamma_\mu(1-\gamma_5)c_\beta]~,\nonumber\\ 
c_4(m_b) & = & -0.033~,~~~O_4  =  [\bar s_\alpha\gamma^\mu(1-\gamma_5) 
b_\beta][\bar c_\beta\gamma_\mu(1-\gamma_5)c_\alpha]~,\nonumber\\ 
c_5(m_b) & = & 0.0096~,~~~~O_5 = [\bar s_\alpha\gamma^\mu(1-\gamma_5) 
b_\alpha][\bar c_\beta\gamma_\mu(1+\gamma_5)c_\beta]~,\nonumber\\ 
c_6(m_b) & = & -0.040~,~~~O_6 = [\bar s_\alpha\gamma^\mu(1-\gamma_5) 
b_\beta][\bar c_\beta\gamma_\mu(1+\gamma_5)c_\alpha]~. 
\eea 
Here $\alpha, \beta$ are color indices. The Wilson coefficients $c_i 
(i = 3 - 6)$ have been calculated in the next-to-leading logarithmic 
approximation (NLL). The second vertex in this diagram is described by 
\beq\label{cslnu} 
H^{\bar c s\to \ell^-\bar\nu_{\ell}}_{\rm eff} = 
\frac{G_F}{\s}V_{cs}[\bar c\gamma^\mu(1-\gamma_5)s] 
[\bar\ell\gamma_\mu(1-\gamma_5)\nu_{\ell}]~. 
\eeq 
 
We are interested in the imaginary part of the amplitude involving the $\bar c s$  
loop illustrated in Fig.\ \ref{fig:loop}.  Contributions of the $(V-A)(V+A)$ operators 
$O_5$ and  
$O_6$ are all proportional to $m_s$ and some are also proportional to $m_\ell$. 
[See Eqs.~(\ref{Tr}) and (\ref{Tmn}).] 
These contributions will be neglected. We consider  
the dominant terms from $O_3$ and $O_4$. 
After Fierz rearrangement of these terms the second order loop amplitude is given by  
\beq  
M_1 = \frac{G_F^2}{2}(c_3 + N_C c_4)V_{tb}V^*_{ts}V_{cs}[\bar c \gamma_\mu  
(1 - \gamma_5) b] T^{\mu \nu} [\bar \ell \gamma_\nu (1 - \gamma_5) \nu_\ell]~,  
\eeq  
where  
\beq\label{Tr}  
T^{\mu \nu} \equiv - \int \frac{d^4k}{(2 \pi)^4} {\rm Tr}\left[ \gamma^\mu  
(1 - \gamma_5) \frac{i(\slk + \slq + m_s)}{(k+q)^2 - m_s^2} \gamma^\nu(1 -  
\gamma_5) \frac{i(\slk + m_c)}{k^2 - m_c^2} \right]~.  
\eeq  
The minus sign takes account of the closed fermion loop.  The most general  
form of $T^{\mu \nu}$ is  
\beq\label{Tmn}  
T^{\mu \nu} = T_1 g^{\mu \nu} + T_2 q^\mu q^\nu~.  
\eeq  
We shall be interested in the $T_1$ part, neglecting the $T_2$ contribution  
to $M_1$ which is proportional to $m_\ell$.  Contracting  
$T^{\mu \nu}$ with $g_{\mu \nu}$ and $q_\mu q_\nu$ we obtain  
\beq  
g_{\mu \nu} T^{\mu \nu} = 4 T_1 + q^2 T_2~,~~  
q_\mu q_\nu T^{\mu \nu} = q^2 T_1 + (q^2)^2 T_2~,{~~\rm so}  
\eeq  
\beq  
T_1 = \frac{1}{3}\left( g_{\mu \nu} T^{\mu \nu} - \frac{q_\mu q_\nu}{q^2}  
T^{\mu \nu} \right)~.  
\eeq  
Performing the appropriate traces, one finds  
\beq \label{eqn:T1}  
T_1 = \frac{1}{6} \int \frac{d^4k}{\pi^4} \left( -\frac{2(q \cdot k)^2}  
{q^2} - k^2 - 3 k \cdot q \right) \left[(k+q)^2 - m_s^2 \right]^{-1} \left[  
k^2 -m_c^2 \right]^{-1}~.  
\eeq 
 
To take twice the absorptive part \cite{PS}, we put the internal propagators 
on the mass shell, replacing  
\beq  
[(k+q)^2 - m_s^2]^{-1} \Rightarrow -2 \pi i \delta[(k+q)^2 - m_s^2]~,  
~~ [k^2 - m_c^2]^{-1} \Rightarrow -2 \pi i \delta[k^2 - m_c^2]~.  
\eeq  
On-shell, we have $k^2 = m_c^2$ and $(k+q)^2 = m_s^2$, implying $2 k \cdot q =  
m_s^2 - m_c^2 - q^2$.  Simplifying by neglecting $m_s$, we find the expression  
in the large brackets in Eq.\ (\ref{eqn:T1}) reduces to $(1/2q^2)(2q^4 - m_c^2  
q^2 - m_c^4)$. The delta functions reduce the loop integral to an integral over 
two-body phase space:  
\bea  
&& \int d^4 k \delta[(k+q)^2 -  m_s^2] \delta[k^2 - m_c^2]  =     
\int d^4 k \int d^4 p~\delta^4(q + k - p) \delta(p^2 - m_s^2)  
\delta(k^2 - m_c^2) \nonumber \\  
&&~~~~~~~~~~~~~~~~~~~~~~~~~~~~~~~~~~~~~ =  
 \int \frac{d^3 k}{2 E_k} \int \frac{d^3 p}{2 E_p}  
\delta^4(q + k - p)  =  (2 \pi)^2 \int d^2 (ps)\,,  
\eea  
where the two-body phase space integral in the limit $m_s = 0$ is  
\beq  
\int d^2 (ps) = \frac{q^2- m_c^2}{8 \pi q^2}~;~~
\int d^n (ps) \equiv (2\pi)^4 \left[ \prod_{i}^n\int \frac{d^3p_i}
{2E_i (2\pi)^3} \right] \delta^4 (P_{\rm final} - P_{\rm initial})~.
\eeq  
Putting the pieces together, we find  
\beq 
T_{1({\rm abs})} = - \frac{q^2 - m_c^2}{12 \pi q^2} \left( \frac{2 q^4  
 - m_c^2 q^2 - m_c^4}{q^2} \right) = - \frac{(q^2-m_c^2)^2}{12 \pi (q^2)^2}  
(2 q^2 + m_c^2)~, 
\eeq  
\beq  
M_{1({\rm abs})} = \frac{G_F^2}{2}(c_3 + N_C c_4) V_{tb}V^*_{ts}V_{cs}  
T_{1({\rm abs})}(q^2) [\bar c \gamma^\mu(1-\gamma_5) b][\bar \ell \gamma_\mu  
(1 - \gamma_5) \nu_\ell]~.  
\eeq  
 
Given the tree level amplitude $M_0$ in Eq.~(\ref{tree}), the expression for
the semileptonic asymmetry then becomes, performing the three-body phase space
integrals,
\beq  
A_{sl} = \frac{2 \int d^3(ps) {\rm Im}(M_0 M^\dag_{1{(\rm abs)}})}  
{\int d^3(ps) |M_0|^2} = -(c_3 + N_C c_4) \frac{{\rm Im} [V^*_{tb} V_{ts}  
V_{cb} V^*_{cs}]}{|V_{cb}|^2} \frac{G_F}{\sqrt{2}} \frac{m_b^2}{6 \pi} R~,  
\eeq  
where 
\beq \label{eqn:R}   
R \equiv \frac{\int dz F(z) G(z)}{\int dz F(z)}~, 
\eeq 
\beq\label{eqn:FG} 
F(z) \equiv  w(z)(z w_1 + w_2 w_3) ~,~~~~ 
G(z) \equiv \left(1-r_c/z\right)^2 (2z + r_c)~. 
\eeq 
Here $r_c \equiv m_c^2/m_b^2,  z \equiv q^2/m_b^2$, $w_1 \equiv 1 + r_c - z$,  
$w_2 \equiv 1 - r_c + z$, $w_3 \equiv 1 - r_c - z,$
and $w(z) \equiv 
(w_2^2 - 4z)^{1/2}$.  The upper limit of integration in both integrals in the  
numerator and denominator of $R$ is  
$z_{\rm max} = (1 - \sqrt{r_c})^2$.  The lower limit of the integral in the  
numerator is $z_{\rm min} = r_c$ (in the limit that the strange quark  
mass may be neglected), while the lower limit in the denominator is zero in the
limit of vanishing lepton mass. Thus the denominator obtains the well-known  
expression for the kinematic factor in $b\to c\ell \bar\nu_{\ell}$, 
\beq  
2 \int_0^{(1 - \sqrt{r_c})^2} dz F(z) = 1 - 8r_c + 8 r_c^3 - r_c^4 + 12 r_c^2  
\ln (1/r_c)~.  
\eeq 
 
For $m_c$ and $m_b$ we use constituent masses, which we expect to simulate QCD 
effects to a certain degree.  These masses were found in Ref.\ 
\cite{Quigg:1981} to reproduce charmonium and bottomonium spectra for a 
rather wide range as long as $m_b - m_c$ lay within a narrower range.  The 
masses considered there are summarized in Table \ref{tab:masses}, together 
with the corresponding values of $r_c$, $R$, and $R m_b^2$ (in which the 
dependence of $R$ on $m_b$ is partially compensated).  The uncertainty on 
$R m_b^2$ is approximately 28\%, which we carry in our final estimate of the 
asymmetry. 
 
Using CKM fits~\cite{Charles:2004jd}, 
\beq\label{eqn:SMphase} 
\frac{{\rm Im}\left(V_{tb}^\star V_{ts} V_{cb} V_{cs}^\star \right)}{\mid V_{cb} \mid^2} 
\approx {\rm Arg}\left(V_{tb}^\star V_{ts} V_{cb} V_{cs}^\star \right) \equiv -\beta_s = -0.018~, 
\eeq 
we then find 
\beq 
A_{s \ell} = (-3.2 \pm 0.9) \times 10^{-9}~. 
\eeq 
\bigskip 
 
\begin{table} 
\caption{Range of constituent-quark masses providing adequate descriptions 
on charmonium and bottomonium spectra
\cite{Quigg:1981}. 
\label{tab:masses}} 
\begin{center} 
\begin{tabular}{c c c c c} \hline \hline 
$m_b$ & $m_c$ &   $r_c =$     &  $R$  & $m_b^2 R$ \\ 
(GeV) & (GeV) & $m_c^2/m_b^2$ &     & (GeV$^2$) \\ \hline 
  4.5 & 1.082 &    0.058      & 0.296 &   5.99    \\ 
 4.75 & 1.359 &    0.082      & 0.207 &    4.67   \\ 
  5.0 & 1.626 &    0.106      & 0.136 &     3.39   \\ \hline \hline 
\end{tabular} 
\end{center} 
\end{table} 
\section{Direct asymmetry beyond the Standard Model} 
A crude estimate for a maximum asymmetry based on dimensional arguments was  
given in Eq.~(\ref{eqn:crude bound}). It applies to a generic new physics 
contribution to the asymmetry occurring at one-loop order. Here we wish to be more 
concrete by considering a specific model leading to CP-violation in $B \to X 
\ell^+\nu_{\ell}$ at one loop, without involving suppression factors which 
occur in the CKM framework. 
 
An example that falls into this category 
is a Left-Right symmetric model~\cite{LRSM1},  
in which the interaction of 
two charged vector-bosons $W_1$ and $W_2$ is 
given by 
\bea\label{eqn:LRSM} 
{\cal L}_W^{LR} =&& \frac{g}{\sqrt{2}}  \bar u_i \left( 
\cos\xi V^L_{ij} \gamma^\mu P_L - e^{i \omega} \sin\xi V^R_{ij}  
\gamma^\mu P_R \right) d_j W_{1 \mu} + \nonumber \\ 
&& \frac{g}{\sqrt{2}}  \bar u_i \left( 
e^{-i \omega} \sin\xi V^L_{ij} \gamma^\mu P_L +   
\cos\xi V^R_{ij} \gamma^\mu P_R 
\right) d_j W_{2 \mu} + {\rm H.c.}~. 
\eea 
Here $P_{L,R} \equiv (1 \pm \gamma_5)/2$ while $V^{L,R}$ are CKM-like matrices 
for left (right)-handed quark fields. The angle $\xi$ is a (small) mixing  
angle between the two charged vector-bosons $W_1$ and $W_2$ and $\omega$  
is a new CP phase related to this mixing. The light mass eigenstate is 
identified with the Standard Model (SM) gauge boson $W_L$, $W_1 \sim W_L$ with 
$M_1 = M_W$.   
 
The interaction (\ref{eqn:LRSM}) contributes to the asymmetry in $B\to X\ell^+ 
\nu_{\ell}$ at one-loop order.  We consider a $W_1$ exchange diagram as in 
Fig.~\ref{fig:tree}, with self-energy insertions on the $W_1$ line of $\bar
cs$ and $\bar ud$ quark loops and $\ell^{\prime}\nu_{\bar\ell^\prime}~
(\ell^\prime = [e,\mu]\ne \ell)$ and $\tau\bar\nu_\tau$ leptonic loops. Elastic
weak rescattering  from an intermediate $\ell\bar\nu$ state is not included for
consistency with CPT~\cite{Gerard:1988jj}.
Loop amplitudes involving $W_2$ exchange or mixed $W_1$-$W_2$ 
exchanges are expected to be smaller 
because a suppression factor $\sin\xi$ is replaced in these amplitudes by  
$M^2_1/M^2_2$.  One obtains $M_2 > 1.6$ TeV if $V_L = V_R$
\cite{Beall:1981ze} (a later estimate gave 1.4 TeV \cite{Langacker:1989xa}) but 
this bound can be relaxed if the left-hand and right-hand quark
mixings are different \cite{Langacker:1989xa}.
For the case $V_L=V_R$, which we consider below, the $W_1$ exchange diagram 
with self energy insertions dominates over these other contributions. 

The absorptive part of the dominant one-loop amplitude is 
\bea
M_{1({\rm abs})}^{LR} = -\frac{G_F^2}{2} N_C \sin\xi \cos^3\xi V_{cb}^R e^{i\omega} 
\cdot T_{1({\rm abs})}^{LR} (q^2) \cdot 
[\bar c \gamma^\mu(1+\gamma_5) b][\bar \ell \gamma_\mu 
(1 - \gamma_5) \nu_\ell]~, 
\eea 
with 
\begin{eqnarray} 
T_{1({\rm abs})}^{LR} (q^2) = 
\mid V_{cs}^L \mid^2 T_{1({\rm abs})}^{cs} + 
\mid V_{ud}^L \mid^2 T_{1({\rm abs})}^{ud}  +
(1/N_C)\left (T_{1({\rm abs})}^{\ell^\prime\nu_{\ell^\prime}} + 
T_{1({\rm abs})}^{\tau\nu_\tau}\right )~.
\end{eqnarray} 
Here $T_{1({\rm abs})}^{cs}(q^2)$ is the same as in the SM, $T_{1({\rm abs})}^{cs} 
\equiv T_{1({\rm abs})}$, while $T_{1({\rm abs})}^{ud}$,  
$T_{1({\rm abs})}^{\ell^\prime\nu_{\ell^\prime}}$ and  $T_{1({\rm abs})}^{\tau\nu_\tau}$ 
correspond to a  $\bar ud$ loop and to the two leptonic loops. 
Using the notations of Eq.~(\ref{eqn:FG}) and the approximation
$m_u=m_d= m_{\ell^\prime}  \simeq 0$, $m_\tau \simeq m_c$, one has
\begin{eqnarray}
T_{1({\rm abs})}^{\tau\nu_\tau}(z) \simeq T_{1({\rm abs})}^{cs}(z) = 
-\frac{m_b^2}{12 \pi} G(z) ~,~~ 
T_{1({\rm abs})}^{\ell^\prime\nu_{\ell^\prime}} \simeq T_{1({\rm abs})}^{ud}(z) = 
-\frac{m_b^2}{6 \pi} z ~. 
\end{eqnarray} 

Comparing the contribution of this amplitude to the asymmetry with the  
asymmetry calculated in the SM, we obtain a ratio 
\bea\label{eqn:AsymRatio} 
\frac{A_{s \ell}^{LR}}{A_{s \ell}^{SM}} = 
8 \sin\xi \cos^3\xi\,\frac{m_c}{m_b}\, \frac{N_C}{c_3+N_C c_4} 
\frac{I^{LR}}{I^{SM}} \,\frac{R^{LR}}{R^{SM}}~. 
\eea 
This ratio involves two enhancement factors following from a suppression 
which occurs in the the Standard Model but not in its Left-Right symmetric extension.
The first factor, $N_C/(c_3+N_C c_4) = -34$, originates in a loop suppression of
the Wilson coefficients for penguin operators in (\ref{Heff}). A second potential 
enhancement is due to the ratio of weak phase factors $I^{LR}/I^{SM}$, where 
\begin{eqnarray} 
I^{SM} \equiv \frac{{\rm Im} [V^{L*}_{tb} V_{ts}^L V_{cb}^L V^{L*}_{cs}]}{|V_{cb}^L|^2} ~,~~~ 
I^{LR} \equiv \frac{{\rm Im} [V_{cb}^L V_{cb}^{R *} e^{-i\omega}]}{|V_{cb}^L|^2}~. 
\end{eqnarray} 
While this factor is $-0.018$ in the SM, it may be of order one in the LR model
if $V^R_{cb}=V^L_{cb}$ and if $ \omega$ is large. 
The last ratio in (\ref{eqn:AsymRatio}) depends on quark couplings and on 
phase space.  Neglecting $u, d$ and $s$ quark masses
and setting $|V_{cs}^L|^2 = | V_{ud}^L |^2 = 1$, it is given by 
\beq 
\frac{R^{LR}}{R^{SM}} = \frac{\int dz z w(z) G(z) +  
2 \int dz z^2 w(z) } {\int dz F(z) G(z)}~. 
\eeq 
The upper limit of integration in the three integrals is $z_{\rm max} = (1- \sqrt{r_c})^2$ 
as in (\ref{eqn:R}). 
The lower limit of the first integral in the numerator and the one in the denominator is  
$z_{\rm min} = r_c$, while that of the second integral in the numerator is zero.  
Taking $r_c = 0.082$ as a central value in Table \ref{tab:masses},  
one finds $R^{LR}/R^{SM} = 0.93$. 
 
Combining all factors and assuming that CP-violation 
in semileptonic $B$ decays is dominated by 
the phase $\omega$, one obtains 
\bea 
\left| \frac{A_{s \ell}^{LR}}{A_{s \ell}^{SM}} \right| \simeq 4\times 10^3  
\frac{|V_{cb}^{R}|}{|V_{cb}^{L}|} |\sin\xi\,\sin\omega| ~. 
\eea 
A recent study of phenomenological
constraints on right-handed quark currents obtains an upper bound on a $b\to 
c$ right-handed coupling of several percent 
relative to a left-handed coupling~\cite{Buras:2010pz}, 
$|(V^R_{cb}/V^L_{cb})\tan\xi\cos\omega|$ $= (2.5 \pm 2.5)\times 10^{-2}$. 
A comparable upper bound may be obtained on 
$|(V^R_{cb}/V^L_{cb})\tan\xi\sin\omega|$ from a recent measurement of CP
asymmetry in $B^+\to J/\psi K^+$~\cite{Sakai:2010gj}, 
$A_{\rm CP}(B^+\to J/\psi K^+) =
[-0.76 \pm 0.50~({\rm stat}) \pm 0.22~({\rm syst})]\times 10^{-2}$.
This bound requires assuming that a final state interaction phase difference 
between two interfering $B \to J/\psi K$ hadronic amplitudes, 
for tree-level $(V-A)(V-A)$ and $(V+A)(V-A)$ $b\to c\bar c s$ 
transitions, is not small. Thus, 
the asymmetry in the Left-Right symmetric model may be at most two  
orders of magnitude larger than in the Standard Model.  

\section{Wrong-sign leptons without neutral $B_{(s)}$ mixing}
In Section II we have assumed $A(\bar B^0_k \to X \ell^+ \nu_{\ell}) = 
A(B^0_k \to X \ell^- \bar\nu_{\ell}) = 0$, 
neglecting second order weak contributions to these two processes 
which occur in the CKM framework leading to ``wrong-sign" leptons without 
$B^0_k$-$\bar B^0_k$ 
mixing. Interference between these second order contributions and first order 
tree amplitudes for $\bar B^0_k \to X \ell^-\bar\nu_{\ell}$ and  
$B^0_k \to X \ell^+ \nu_{\ell}$ leads to additional time-dependent 
terms in Eqs.~(\ref{dGamma/dt}) of the forms 
$e^{-\Gamma_kt}\sinh(\Delta\Gamma_kt/2)$ and  
$e^{-\Gamma_kt}\sin(\Delta m_kt)$. 
Second order amplitudes for  $B^0_k \to X \ell^-\bar\nu_{\ell}$ have been 
discussed in Ref.~\cite{Dass:1996wi} without estimating their magnitudes. 
For completeness, as we have studied tiny CP 
asymmetries from second order amplitudes, we will estimate the ratio 
of second order amplitudes for ``wrong-sign" leptons and first order
amplitudes for ``right-sign" leptons, showing that this ratio is negligibly small. 
 
\begin{figure}[h] 
\begin{center} 
\includegraphics[width=0.45\textwidth]{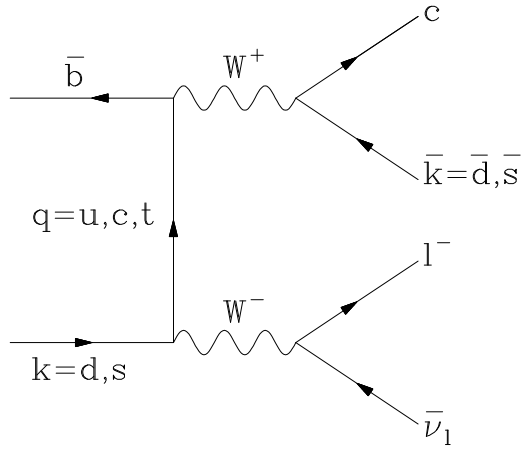} 
\end{center} 
\caption{Second order quark diagrams contributing to $B^0_k \to X\ell^-\bar\nu_{\ell}$. 
\label{fig:ww}} 
\end{figure}

Second order amplitudes for $B^0_k \to X \ell^- \bar\nu_{\ell}$ are described 
by diagrams plotted in Fig.\,\ref{fig:ww}, 
in which both the $\bar b$ quark and the spectator $k$ quark undergo weak 
decays into $\bar q c \bar k$ and $q\ell^-\bar\nu_{\ell}$, respectively, by 
exchanging $q = u, c, t$ quarks. 
These second order amplitudes, involving CKM factors $V^*_{qb}V_{qd}V_{cd}$ 
and $V^*_{qb}V_{qs}V_{cs}$ in $B^0$ and $B_s$ decays, lead to final hadronic 
states with quark structures $X=c \bar d$ and $X = c \bar s$, respectively, 
as in first-order amplitudes for $\bar B^0$ and $\bar B_s$ semileptonic decays. 
 
Let us denote the second-order weak amplitude  for $B^0_k$ decay by 
$A^k_{\ell} \equiv A(B^0_k \to X \ell^- \bar\nu_{\ell})$. 
 We wish to estimate the ratios of semileptonic rates 
\beq \label{eqn:rat} 
R_k = \left| \frac{A^k_{\ell}}{\bar A^k_{\ell}} \right|^2 \frac{\Phi_2}{\Phi_1}~, 
\eeq 
where $\Phi_{1,2}$ are the appropriate phase-space factors for the first-order 
and second-order processes.  By neglecting the effect of the 
spectator quark in $\bar A^k_{\ell}$ we are treating the first-order process as 
leading to a three fermion final state, while the second order diagram 
illustrated in  Fig.\,\ref{fig:ww} involves four fermions in the final state. 
A naive dimensional analysis then leads to 
\beq \label{eqn:phsp} 
\frac{\Phi_2}{\Phi_1} = \frac{(m_b-m_c)^2}{16 \pi^2}~.
\eeq 

 The ratio $A^k_{\ell}/\bar A^k_{\ell}$ must involve a factor of $G_F f_B $ which has 
a suitable dimension. The momentum passing through the propagator 
of the fermion $q = u,c,t$ is of order $m_b$, and kinematic factors of the 
same order will cancel it for $q = u,c$, while the contribution of $q=t$ is 
highly suppressed by the heavy $t$-quark mass.  The corresponding CKM factors 
are $V^*_{ub}V_{us}V_{cs} \sim {\cal O}(\lambda^4)~(q=u), 
V^*_{cb}V_{cs}V_{cs}\sim {\cal O}(\lambda^2)~(q=c)$ and 
$V^*_{ub}V_{ud}V_{cd}\sim {\cal O}(\lambda^4)~(q=u), 
V^*_{cb}V_{cd}V_{cd}\sim {\cal O}(\lambda^4)~(q=c)$ for $B_s$ and $B^0$ 
decays, respectively, 
where $\lambda = 0.23$. 
Thus, the $c$ quark dominates $A^s_{\ell}$ with a CKM factor 
comparable to $V_{cb}$ governing $\bar A^s_{\ell}$, while the contributions of the 
$u$ and $c$ quarks in $A^d_{\ell}$ are comparable to each other and are suppressed 
by  $\lambda^2\simeq 1/20$ relative to the CKM factor in $\bar A^d_{\ell}$. 
Taking $m_b-m_c = 3.4$ GeV (see Table \ref{tab:masses} 
above for values of $m_b$ and $m_c$)  
and $f_B = 230$ MeV, one then finds 
\bea 
R_s & \sim & \frac{(m_b-m_c)^2}{16 \pi^2} (f_B G_F)^2 \simeq 5.3 \times 10^{-13}~,\\ 
R_d & \sim & (1/20)^2 R_s \simeq 1.3 \times 10^{-15}~. 
\eea
The corresponding ratios  of square roots,  $\sim 0.7\times 10^{-6}$ and 
$\sim 0.4\times 10^{-7}$, 
characterize coefficients of additional time-dependent 
terms of forms $e^{-\Gamma_kt}\sinh(\Delta\Gamma_kt/2)$ and  
$e^{-\Gamma_kt}\sin(\Delta m_kt)$ in Eqs.~(\ref{dGamma/dt}) for $B_s$ and $B^0$ 
which may be safely neglected. 
 
\section{Conclusion}  
Inclusive semileptonic $B$ and $B_s$ decays are shown to have a small 
non-zero direct CP asymmetry in the Standard Model as a result of 
interference of first-order- and second-order-weak processes. 
This stands in contrast with statements made in two textbooks on CP 
violation~\cite{Branco_p58,Big_p186}.
Taking a range of effective quark masses and estimating the asymmetry in 
$b \to c \ell \bar \nu_\ell$ as due to 
weak rescattering from the intermediate state
in $b \to_{\hskip -5mm{\rm peng}} c \bar c s$,  we have found 
\beq
A_{sl} = (-3.2 \pm 0.9) \times 10^{-9}~. 
\eeq

A dimensional argument leads to a model-independent upper bound on
$A_{sl}$ which is three orders of magnitude larger, while an  
extension of the Standard Model to a left-right-symmetric variant can 
increase the above asymmetry by at most two orders of magnitude.
These values are far smaller than a value of about $-1\%$ recently 
reported by the D0 Collaboration \cite{Abazov:2010hv}, which may 
still be associated with a new source of CP violation in neutral $B$ 
meson mixing. 

As a by-product of second-order-weak amplitudes, we have estimated 
their efffect on wrong-sign leptons in direct decays of  neutral $B$ mesons
and found it to be much below any reasonable sensitivity. 
 
\section*{Acknowledgments} 
 
M. G. is grateful to Steven and Priscilla Kersten and the Enrico Fermi 
Institute at the University of Chicago for their kind and generous hospitality. 
We thank D. Tonelli for his interest in this question
and B. Bhattacharya for discussions. 
This work was supported in part by the United States Department of Energy 
through Grant No.\ DE FG02 90ER40560.

\end{document}